\documentstyle[epsfig]{aipproc}
\def\slr{\tilde{l}_R}
\def\sll{\tilde{l}_L}
\def\lsp{{\tilde{\chi}}^0_1}
\def\neuttwo{{\tilde{\chi}}^0_2}
\def\chargone{{\tilde{\chi}}^{\pm}_1}
\def\chargtwo{{\tilde{\chi}}^{\pm}_2}
\def\selright{\tilde{e}_R}
\def\selleft{\tilde{e}_L}
\def\smuright{\tilde{\mu}_R}
\def\smuleft{\tilde{\mu}_L}
\def\snu{{\tilde{\nu}}_L}
\def\gluino{\tilde{g}}
\def\stopone{{\tilde{t}}_1}
\def\sbotone{{\tilde{b}}_1}
\def\higgsA{H_A}
\def\lhiggs{h_0}
\def\tanb{{{\rm tan}\,\beta}}
\def\msmu{M_{\tilde{\mu}}}
\def\mlsp{M_{\rm LSP}}
\def\emax{E_{\mu}^{\rm max}}
\def\emin{E_{\mu}^{\rm min}}
\setbox3=\hbox to0.5in{}
\setbox4=\hbox to0.5in{}

\begin{document}
\title{Sparticle masses from kinematic\\ fitting at a muon collider
\thanks{Talk presented at the 4th International Conference on the
Physics Potential and Development of $\mu^+\mu^-$ Colliders,
San Francisco, 10-12 December, 1997.}
}
\author{Joseph D. Lykken$^*$}
\address{$^*$Theoretical Physics Department\\
Fermi National Accelerator Laboratory\\
P.O. Box 500\\
Batavia, IL 60510}

\maketitle

\begin{abstract}
Three case studies are presented of slepton pair
production followed by two-body or quasi-two-body decays at a muon collider.
Precision mass measurements are possible using a variety
of kinematic fitting methods. Standard Model and supersymmetric
backgrounds are easily controlled by kinematic cuts.
In all three cases it appears that detector resolutions,
not backgrounds or statistics, will dominate the final
error bars. Polarized beams are not necessary
to control SM backgrounds. However, without polarization
it may be difficult in some cases to disentangle $\slr$
from $\sll$ signals. 

\end{abstract}

\subsection*{Introduction}

A muon collider is in principle an excellent machine for precision
studies of weak scale supersymmetry. Depending on $\sqrt{s}$ and
the SUSY mass spectrum, it may be possible to observe pair production of a
half-dozen or more distinct sparticles. For R parity preserving
SUSY, sparticle pair production is kinematically underconstrained,
due to the pair of unmeasured LSP's. However in many cases each
sparticle in the pair has a significant branching fraction for
what is essentially a two-body decay:

\begin{equation}
{\rm sparticle} \rightarrow {\rm LSP} + {\rm particle}\quad,
\end{equation}
where ``particle'' refers to a fully reconstructible Standard
Model particle (e, $\mu$, W, Z, and possibly $\lhiggs$, t),
while the LSP is assumed to be the lightest neutralino
$\lsp$.

In these cases there are a variety of kinematic fitting methods for
extracting sparticle masses. In this talk I will report on
two such methods applied to smuon, selectron, and sneutrino
production at a muon collider. Chargino production is not discussed,
since for light fermionic sparticles the best method for a precision
mass measurement is a threshold scan\cite{lykmberger}. An interesting
challenge for future investigation is the production of 
staus, stops, and the heavier chargino and neutralinos.

Sparticle production at a muon collider is similar in
many respects to sparticle production at an e$^+$e$^-$ machine.
For the present analysis the most important differences are
that the muon collider has (i) much higher energy reach,
(ii) significantly lower advertised luminosity at comparable
energies, (iii) little or no polarization available without
taking a significant hit in luminosity, and (iv) large
detector backgrounds from muon decays. These detector backgrounds
are generally soft, but large fluctuations could cause problems
for precision SUSY measurements. They will also impact on isolation
cuts, determinations of missing $E_T$, and detector resolutions
generally. These problems will be left to future study. 

At a muon collider smuon pairs arise from both s and t channel production;
the s channel production is through a virtual photon or Z, while
the t channel diagram involves the exchange of a neutralino.
The s and t channel contributions interfere destructively, but
this effect will not be important for the examples considered
here, where the t channel production is dominant. Selectron production
proceeds only through the s channel, and is thus suppressed in the
examples. Muon sneutrino production proceeds only through the
t channel, and is thus competitive with smuons.

In both supergravity (sugra) and gauge mediated models,  
the $\slr$'s are lighter than the $\sll$'s.
Independent of the SUSY model, $\slr$'s decay almost 100\%
via a single two-body mode:
\begin{equation}
\smuright\rightarrow\lsp\;\mu \quad ;\qquad
\selright\rightarrow\lsp\;{\rm e}\quad .
\end{equation}
The branching fractions of the $\sll$'s and $\snu$'s
are model dependent. The important decay modes for, e.g.,
$\smuleft$ are:
\begin{eqnarray}
\smuleft \rightarrow &\lsp\;\mu \nonumber \\
&\neuttwo\;\mu \\
&~~~~~~\chargone\;\nu_{\mu} \quad .\nonumber
\end{eqnarray}

\subsection*{Kinematics}

The basic kinematics can be understood by considering
pair production of $\smuright$:
\begin{eqnarray}
\mu^+\mu^- \rightarrow & ~~~~\smuright (p_1)\smuright (p_2)& \nonumber\\
&\smuright (p_1) \rightarrow & \lsp (p_3) \mu (p_4) \nonumber\\
&\smuright (p_2) \rightarrow & \lsp (p_5) \mu (p_6) \quad .
\end{eqnarray}
Each event consists of an acoplanar dimuon pair plus missing $E_T$.
Six measurements are made, i.e., the 3-momenta of the
two muons. The event is characterized by 13 kinematic variables:
the four 3-momenta of the final state plus the common LSP
mass. There are 5 kinematic constraints: one from the assumption
that the two smuons have the same mass, and the rest from
the known initial state 4-momentum.
This leaves 2 undetermined variables
in the event, which we may take as $\msmu$,
$\mlsp$.

\begin{table}
\caption{Sparticle and Higgs spectrum for LHC Point 5, which
corresponds to minimal sugra parameters $m_0$$=$$100$ GeV,
$m_{1/2}$$=$$300$ GeV, $A_0$$=$$0$, $\tanb$$=$$2.1$, and
sgn($\mu$)$=$1.}
\label{lyktable1}

\begin{tabular}{lddld}
   Particle  & Mass (GeV)&\box3& Particle &
   Mass (GeV)\\
\tableline
$\lsp$      & 119 && $\neuttwo$  & 228\\
$\chargone$ & 228 && $\chargtwo$ & 565\\
$\selright$ & 157 && $\selleft$  & 241\\
$\smuright$ & 157 && $\smuleft$  & 241\\
$\snu$      & 232 && $\gluino$   & 754\\
$\stopone$  & 448 && $\sbotone$  & 604\\
$\lhiggs$   &  94 && $\higgsA$   & 657\\
\end{tabular}
\end{table}

The kinematic endpoint method arises from the following
expression for the energy of each muon as measured in the
rest frame of its parent smuon:
\begin{equation}
E_{\mu}^0 = {\msmu^2 - \mlsp^2\over 2\msmu} \quad ,
\end{equation}
where we are neglecting the muon mass. The maximum and minimum
boosts from this frame to the lab frame then provides us with
two kinematic endpoints $\emax$, $\emin$, in the muon energy spectrum.
A precision measurement of both endpoints allows us to extract
both $\msmu$ and $\mlsp$. Note that this method requires good statistics
to be useful, and does not take advantage of all the kinematic information
in the event.

Another kinematic method, developed by Feng and Finnell for
e$^+$e$^-$ studies, extracts $\msmu$ assuming a precision
value of $\mlsp$ is already known from other sources and thus
can be used as an input. This method starts with the relation
\begin{equation}
\msmu^2 = {1\over 4}s - \vert\vec{p}_3\vert^2 - \vert\vec{p}_4\vert^2
- 2\vert\vec{p}_3\vert\vert\vec{p}_4\vert{\rm cos}\,\theta_{34} \quad .
\end{equation}
For given input value of $\mlsp$, the only unknown on the
right hand side is $\theta_{34}$, the angle between the
3-vectors $\vec{p}_3$ and $\vec{p}_4$. This angle is then
estimated, event by event, by a certain function of measured variables.
This function has the property that the error of the estimate
goes to zero in the limit that the two LSP's are back-to-back
in the lab frame. For $\sqrt{s}/2 \gg \msmu \gg \mlsp$, the
mass estimates peak strongly around the true value, and precise
results are possible even for rather sparse data.

A third kinematic method, which is currently under
investigation, involves adapting the likelihood methods
developed for extracting the top quark mass from the
dilepton channel. This method is also well-suited to
sparser data sets.

\subsection*{Simulations}

Simulations were performed using PYTHIA v6.1
\cite{lykpythia} coupled to
the ATLFAST v1.25 
\cite{lykatlfast} fast detector simulator.
Note that the small differences in the sparticle
spectra produced by PYTHIA and ISAJET\cite{lykisajet} make a
difference for the analysis done here.
The ATLFAST
defaults were used for lepton isolation and jet reconstruction.
Smearing was not included, and no attempt was made to
include detector backgrounds. Thus ``precision'' here
refers only to statistics and to SUSY signal versus
Standard Model (SM) backgrounds and SUSY backgrounds.

\begin{figure}[b!] 
\centerline{\epsfig{file=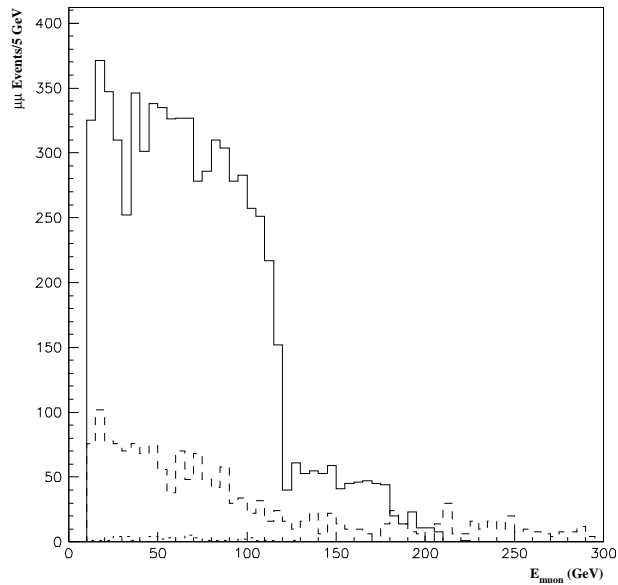,height=3.5in,width=3.5in}}
\vspace{10pt}
\caption{Dimuon production after cuts, 20 fb$^{-1}$ at
$\sqrt{s}$$=$$600$ GeV for LHC point 5. The solid line is the
total smuon signal. The dashed line is the sum of the
Standard Model backgrounds; the dotted line is the background
from chargino pairs.}
\label{lykfig1}
\end{figure}

\subsection*{Sleptons at LHC point 5}

This first study overlaps with the analysis presented
by Frank Paige at the Fermilab workshop\cite{lykfpaige}.
LHC point 5 is
a mimimal supergravity reference point described in
Table \ref{lyktable1}.
For dimuon and dielectron production at $\sqrt{s}$$=$$600$ GeV,
cuts were imposed similar to those of \cite{lykfpaige}:

\begin{figure}[b!] 
\centerline{\epsfig{file=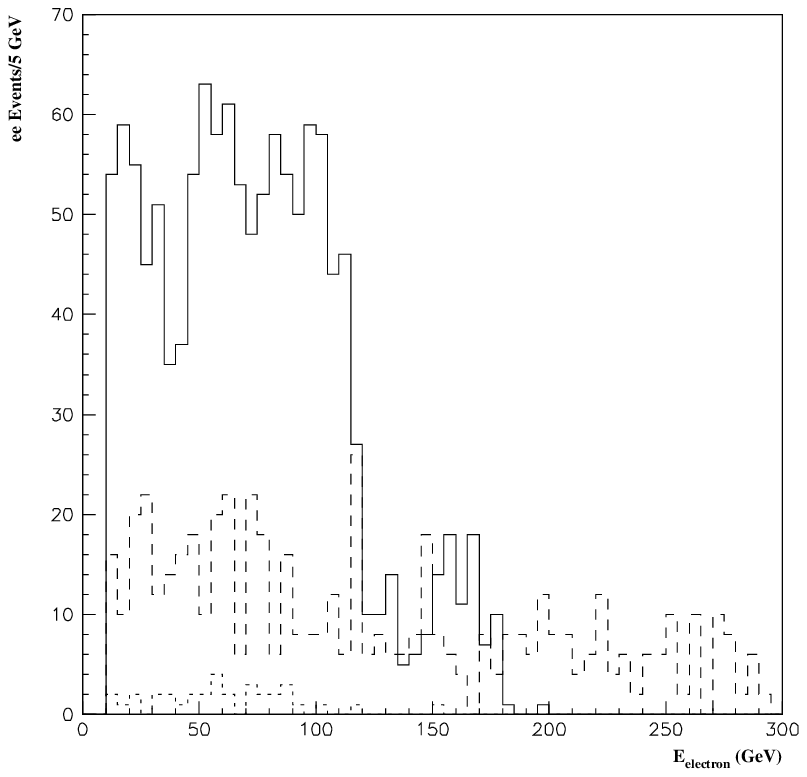,height=3.5in,width=3.5in}}
\vspace{10pt}
\caption{Dielectron production after cuts, 20 fb$^{-1}$ at
$\sqrt{s}$$=$$600$ GeV for LHC point 5. The solid line is the
total selectron signal. The dashed line is the sum of the
Standard Model backgrounds; the dotted line is the background
from chargino pairs.}
\label{lykfig2}
\end{figure}

\begin{itemize}
\item Exactly two isolated e or $\mu$ leptons and no jets,
\item $E>10$ GeV and $\vert\eta\vert < 1.3$ for each lepton,
\item $\Delta\phi_{1,2} < 0.9\pi$,
\item $\vert\vec{p}_{T,1}+\vec{p}_{T,2}\vert > 10$ GeV, and
\item missing $E_T >$ 20 GeV.
\end{itemize}
Note that missing $E_T$ signatures are degraded at a muon
collider detector, due to the 20 degree forward and backward
dead cones needed for shielding. This is not a crucial point
for the present analysis, however.

The signal acceptance with these cuts is approximately 40\%.
The cuts are very efficient at eliminating backgrounds. The
simulations included the six most important SM backgrounds;
these are:
\begin{itemize}
\item $W^+W^-$ pair production,
\item $\gamma\mu \rightarrow W\nu_{\mu}$,
\item Drell-Yan,
\item $\gamma\mu \rightarrow Z\mu$,
\item $\gamma\gamma \rightarrow l^+l^-$, and
\item $ZZ$ pairs.
\end{itemize}
The main SUSY background is from chargino pair production,
with both charginos decaying leptonically.

\begin{figure}[b!] 
\centerline{\epsfig{file=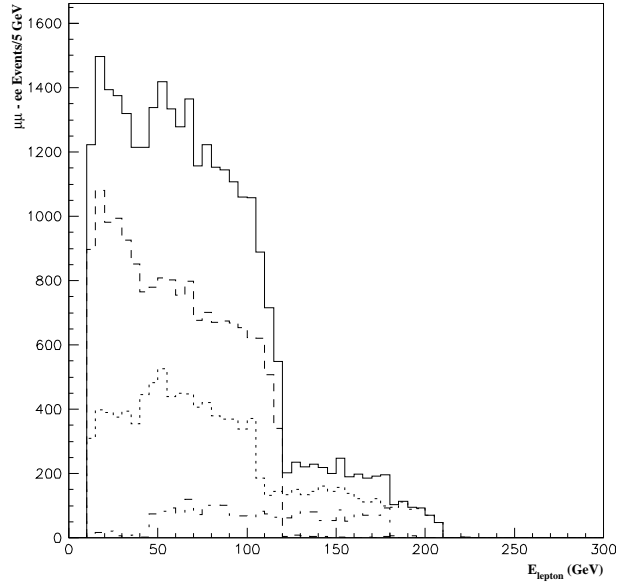,height=3.5in,width=3.5in}}
\vspace{10pt}
\caption{Flavor subtracted slepton signal, 100 fb$^{-1}$ at
$\sqrt{s}$$=$$600$ GeV for LHC point 5. The solid line is the
total smuon + selectron signal. The dashed, dotted, and dot-dashed
lines are from $\slr\slr$, $\slr\sll$$+$$\sll\slr$, and
$\sll\sll$, respectively.}
\label{lykfig3}
\end{figure}

Figures \ref{lykfig1} and \ref{lykfig2} show the dimuon
and dielectron event rates plotted versus muon or electron
energy, with 5 GeV bins.
The SM backgrounds after cuts are rather flat and
encouraging small, even for the dielectron case. The SUSY
background is negligible. 
For 20fb$^{-1}$ of integrated luminosity, Figure \ref{lykfig2}
also reflects the rather poor statistics of selectron production.
This is not surprising given that the total cross section is
only 64 fb. The situation is noticeably better for smuon
production, where the cross section is 400 fb.

Figure \ref{lykfig3} shows the $\mu\mu -$ee flavor subtracted
slepton signal, after cuts, broken down into the its three components:
RR, RL+LR, and LL. The integrated luminosity is 100 fb$^{-1}$
to enhance the statistics. As discussed in \cite{lykfpaige},
this figure shows a rather complicated structure, reflecting
the fact that there are eight distinct kinematic endpoints
affecting the distribution. These are:
\begin{eqnarray}
\slr\slr :~~~&118 {\rm\ GeV},~~~&~9 {\rm\ GeV};\nonumber\\
\slr\sll :~~~&105 {\rm\ GeV},~~~&11 {\rm\ GeV};\nonumber\\
\sll\slr :~~~&208 {\rm\ GeV},~~~&40 {\rm\ GeV};\\
\sll\sll :~~~&181 {\rm\ GeV},~~~&46 {\rm\ GeV}.\nonumber
\end{eqnarray}
Comparing with Figures \ref{lykfig1},\ref{lykfig2}, it
appears that with 20 fb$^{-1}$ and a perfect detector,
one can determine the endpoints at 118, 208, and 181 GeV
to an accuracy of one bin or better. The other endpoints
look very challenging.

\begin{figure}[b!] 
\centerline{\epsfig{file=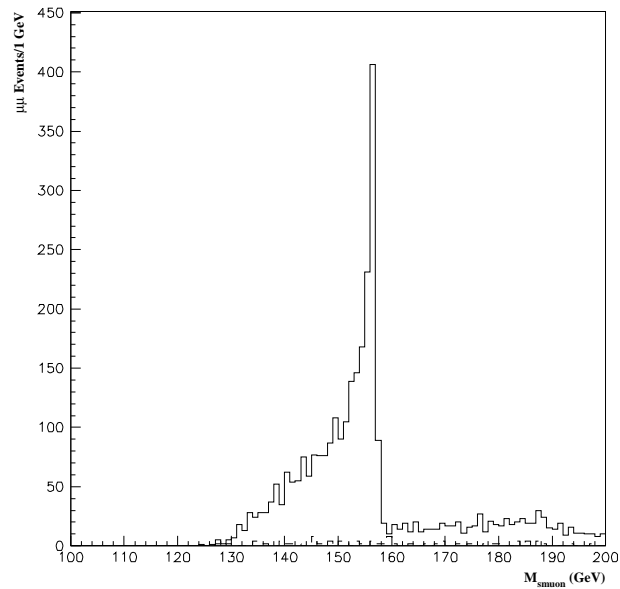,height=3.5in,width=3.5in}}
\vspace{10pt}
\caption{Dimuon production after cuts, 20 fb$^{-1}$ at
$\sqrt{s}$$=$$600$ GeV for LHC point 5. The solid line is the
total smuon signal, plotted versus the Feng-Finnell
estimate for the smuon mass. The dashed line is the sum of the
Standard Model backgrounds.}
\label{lykfig4}
\end{figure}

The situation improves if we include the Feng-Finnell
estimate for the smuon mass. This is shown in
Figure \ref{lykfig4}, plotted with 1 GeV bins.
The SM background shown is completely negligible.
Because of the strong peaking, which actually resembles
a sharp edge,
it is trivial to extract
the $\smuright$ mass with an accuracy of one bin or
better. This assumes that the $\lsp$ mass is already known
to within 1 GeV. Similar results are obtained for the
$\selright$, with somewhat worse statistics.

\subsection*{Heavy sleptons}

The second study is for the heavy sugra point described
in Table \ref{lyktable2}. The results are for dimuon and
dielectron production at $\sqrt{s} = 1400$ GeV, using the
same cuts as in the previous example.

\begin{table}
\caption{Sparticle and Higgs spectrum for the heavy
sugra point, which
corresponds to minimal sugra parameters $m_0$$=$$500$ GeV,
$m_{1/2}$$=$$350$ GeV, $A_0$$=$$0$, $\tanb$$=$$2$, and
sgn($\mu$)$=$-1.}
\label{lyktable2}

\begin{tabular}{lddld}
   Particle  & Mass (GeV)&\box3& Particle &
   Mass (GeV)\\
\tableline
$\lsp$      & 145 && $\neuttwo$  &  290\\
$\chargone$ & 290 && $\chargtwo$ &  809\\
$\selright$ & 519 && $\selleft$  &  561\\
$\smuright$ & 519 && $\smuleft$  &  561\\
$\snu$      & 558 && $\gluino$   &  886\\
$\stopone$  & 597 && $\sbotone$  &  763\\
$\lhiggs$   &  84 && $\higgsA$   & 1083\\
\end{tabular}
\end{table}

\begin{figure}[b!] 
\centerline{\epsfig{file=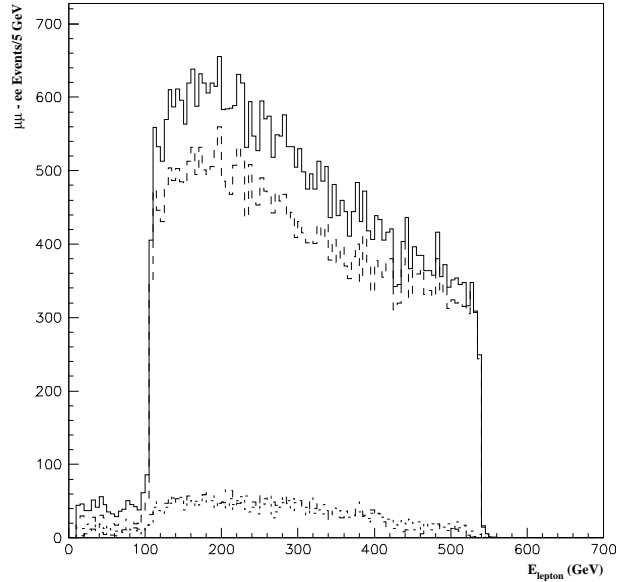,height=3.5in,width=3.5in}}
\vspace{10pt}
\caption{Flavor subtracted slepton signal, 1000 fb$^{-1}$ at
$\sqrt{s}$$=$$1400$ GeV for the heavy sugra point. The solid line is the
total smuon + selectron signal. The dashed, dotted, and dot-dashed
lines are from $\slr\slr$, $\slr\sll$$+$$\sll\slr$, and
$\sll\sll$, respectively.}
\label{lykfig5}
\end{figure}

Figure \ref{lykfig5} shows the flavor subtracted slepton
signal corresponding to 1000 fb$^{-1}$ of integrated luminosity.
Comparing with Figure \ref{lykfig3}, one notes several
differences.
In the present case the signal is completely
dominated by RR production. This is because the branching
fraction for $\smuleft$ or $\selleft$ decay to muon or
electron plus $\lsp$ is only 16\%. At this heavy sugra point,
the $\smuleft$ decays predominantly to either $\chargone\,\nu_{\mu}$
or $\neuttwo\,\mu$. Subsequent decays in these modes are unlikely
to pass the cuts. 

Since RR production now dominates, there are effectively only two
kinematic endpoints: 539 GeV and 106 GeV. Note that the lower
endpoint is now sufficiently large not to be distorted or
hidden by the cuts. Both edges are very sharp in Figure
\ref{lykfig5}. The SM backgrounds after cuts are negligible.
Thus with a perfect detector one could extract the masses of
both $\smuright$ and $\lsp$ with an accuracy better than 5 GeV.

\begin{figure}[b!] 
\centerline{\epsfig{file=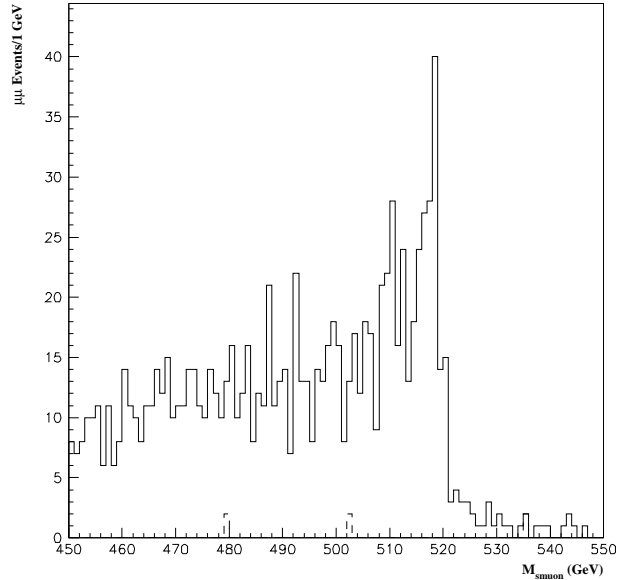,height=3.5in,width=3.5in}}
\vspace{10pt}
\caption{Dimuon production after cuts, 100 fb$^{-1}$ at
$\sqrt{s}$$=$$1400$ GeV for the heavy sugra point. The solid line is the
total smuon signal, plotted versus the Feng-Finnell
estimate for the smuon mass. The dashed line is the sum of the
Standard Model backgrounds.}
\label{lykfig6}
\end{figure}

Figure \ref{lykfig6} shows the Feng-Finnell plot for the
heavy sugra point. The SM backgrounds shown are negligible.
Again we see strong edgelike peaking around the actual $\smuright$
mass of 519 GeV.
It is clearly possible to extract
the mass with an accuracy of one bin or
better. This assumes that the $\lsp$ mass is already known
to within 1 GeV. Similar results are obtained for the
$\selright$, but with poor statistics.

\begin{table}
\caption{Sparticle and Higgs spectrum for the third sugra point, which
corresponds to minimal sugra parameters $m_0$$=$$225$ GeV,
$m_{1/2}$$=$$200$ GeV, $A_0$$=$$0$, $\tanb$$=$$2$, and
sgn($\mu$)$=$1.}
\label{lyktable3}

\begin{tabular}{lddld}
   Particle  & Mass (GeV)&\box3& Particle &
   Mass (GeV)\\
\tableline
$\lsp$      &  77 && $\neuttwo$  & 146\\
$\chargone$ & 144 && $\chargtwo$ & 449\\
$\selright$ & 240 && $\selleft$  & 270\\
$\smuright$ & 240 && $\smuleft$  & 270\\
$\snu$      & 262 && $\gluino$   & 536\\
$\stopone$  & 310 && $\sbotone$  & 450\\
$\lhiggs$   &  88 && $\higgsA$   & 560\\
\end{tabular}
\end{table}

\begin{figure}[b!] 
\centerline{\epsfig{file=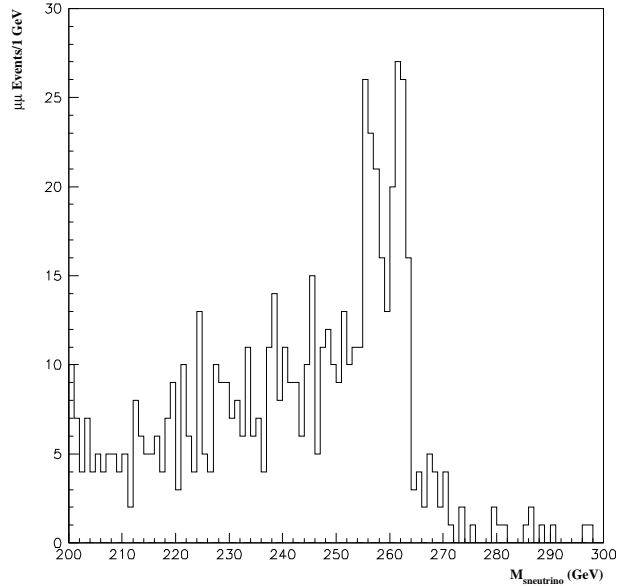,height=3.5in,width=3.5in}}
\vspace{10pt}
\caption{Dimuons plus two or more jets, after cuts, 20 fb$^{-1}$ at
$\sqrt{s}$$=$$800$ GeV for the second sugra point. Shown is the
total SUSY signal, plotted versus the Feng-Finnell
estimate for the sneutrino mass.}
\label{lykfig7}
\end{figure}

\subsection*{Sneutrino pair production}
Muon sneutrino pair production fits our kinematic
scenario, provided that the sneutrino has a substantial
branching fraction to $\chargone\,\mu$. The chargino will
decay predominantly to $\lsp$ plus jets. Thus the signature
is an acoplanar dimuon pair plus missing $E_T$ plus jets.
Note that, in the presence of the R parity violating coupling
LLE, s-channel resonant production of single sneutrinos
may also be possible at a muon collider\cite{lykfeng}.

Here we have studied the sugra point described in Table
\ref{lyktable3}, for production at $\sqrt{s} = 800$ GeV.
The branching fraction for the muon
$\snu$ into $\chargone\,\mu$ is 56\%, while the branching
fraction for $\chargone$ into $\lsp$ plus jets is 65\%.
We will employ the same cuts as previously, except that
we now require two or more reconstructed jets (cone radius
$R = 0.4$).

Figure \ref{lykfig7} shows the Feng-Finnell mass estimate
after cuts plotted in 1 GeV bins. Shown is the total signal from all
SUSY production mechanisms. The SM background after cuts is negligible.
The signal acceptance for muon sneutrino pairs after cuts is
about 4\%. Thus, despite a rather large cross section (over
500 fb) the plot has rather poor statistics. Nevertheless we
again see strong edgelike peaking at the true $\snu$ mass of 262 GeV.

It is interesting to note that a previous study of sneutrino
pair production at e$^+$e$^-$ colliders
\cite{lykbaer} relied on the trilepton plus
missing $E_T$ plus jets channel to kill SM backgrounds. For
our study point this does not appear to be necessary.
Furthermore, our complementary dilepton channel has five times
the rate, before cuts, as the trilepton channel.

\subsection*{Conclusions}

A variety of precision sparticle mass measurements are
possible at a muon collider using kinematic methods such
as those discussed here. Polarized beams are not necessary
to control SM backgrounds. However, without polarization
it may be difficult in some cases to disentangle $\slr$
from $\sll$ signals.

It seems likely that in most cases detector resolutions,
not backgrounds or statistics, will dominate the final
error bars. Thus it will be crucial to perform
simulations with a realistic mock-up of a muon collider
detector.

Adequate statistics for the type of analysis presented
here correspond to integrated luminosities of
at least 20 fb$^{-1}$ for $\sqrt{s}$ in the range
500 to 800 GeV. For heavy sparticles and $\sqrt{s}\gtrsim 1$
TeV, the minimum useful integrated luminosity is about
100 fb$^{-1}$.

This research was
supported by the Fermi National Accelerator Laboratory, which
is operated by Universities Research Association, Inc., under
contract no. DOE-AC02-76CHO3000.

\end{document}